\documentclass[fleqn,twoside]{article}
\usepackage{espcrc2}
\usepackage{graphicx}
\usepackage[figuresright]{rotating}

\newcommand{\AmS}{{\protect\the\textfont2
  	A\kern-.1667em\lower.5ex\hbox{M}\kern-.125emS}}
\title{Baryonic sources using irreducible representations of the 
double-covered octahedral group}
\author{LHP Collaboration:
	S.~Basak\address[umd]{Department of Physics, 
        University of Maryland, College Park, MD 20742, USA},
	R.~Edwards\address[jlab]{Thomas Jefferson National Accelerator 
	Facility, Newport News, VA 23606, USA},
	R.~Fiebig\address{Physics Department, Florida International University,
	Miami, FL 33199, USA},
	G.~T.~Fleming\addressmark[jlab],
	U.~M.~Heller\address{American Physical Society, One Research Road,
	Ridge, NY 11961-9000, USA},
	C.~Morningstar\address{Department of Physics, Carnegie Mellon
	University, Pittsburgh, PA 15213, USA},
	D.~Richards\addressmark[jlab],
	I.~Sato\addressmark[umd]\thanks{Presented by I. Sato}, and
	S.~Wallace\addressmark[umd].}
\begin{document}
\begin{abstract}
Irreducible representations (IRs) of the double-covered octahedral
group are used to construct lattice source and sink operators for three-quark
baryons.  The goal is to achieve a good coupling to 
higher spin states as well as ground states.  Complete sets of local 
and nonlocal straight-link operators are explicitly shown for 
isospin 1/2 and 3/2 baryons.  The orthogonality relations of the 
IR operators are confirmed in a quenched lattice simulation.
\end{abstract}
\maketitle
\section{INTRODUCTION}
The determination of the hadronic spectrum is a crucial goal of lattice QCD.
Lattice simulations should answer such open questions as 
the nature of Roper resonance, 
masses and structures of higher spin states, and explore the existence of 
pentaquarks, hybrids, and other exotic states.  
This paper focuses on the construction of three-quark operators with
zero and nonzero orbital angular momentum.  
The method is applicable to any multi-quark hadron.  
Half-integer spin objects on the lattice can be characterized by
IRs of the spinorial rotation group known as the 
double-covered octahedral group~\cite{Johnson82,Morningstar03}.  
The group has three IRs: $G_1, G_2,$ and $H$ with respective dimensions 2, 2, and 4.  
For states transforming according to $G_1$ possible spins are 
${1\over 2}, {7\over 2}, {9\over 2}, ...$, for $G_2$ spins are 
${5\over 2}, {7\over 2}, ...$, and for $H$ spins are 
${3\over 2}, {5\over 2}, {7\over 2}, {9\over 2}, ...$.

\section{LOCAL OPERATORS}
Assuming that up and down quark masses are degenerate,
a color-neutral nucleon source operator formed from three quarks is
\begin{equation}
\overline N^{\rho_1 \rho_2 \rho_3}_{s_1 s_2 s_3} = 2^{ -{1\over 2}} \epsilon_{abc} 
(\overline u^{\rho_1 a}_{s_1} \overline d^{\rho_2 b}_{s_2} - 
 \overline d^{\rho_1 a}_{s_1} \overline u^{\rho_2 b}_{s_2} ) 
 \overline u^{\rho_3 c}_{s_3} ,
\label{nucleon1}
\end{equation}
where all quark fields are located at $({\bf x}, t)$.
Color indices are labeled by 
$a, b,$ and $c$, and spinor indices are written in terms of 
two-component $\rho$-spin and two-component $s$-spin 
($\rho$-spin denotes upper/lower component of Dirac spinor and $s$-spin 
denotes up/down ordinary spin).  
The $\rho$-spin is the eigenvalue of intrinsic parity: 
$+/-$ for upper/lower components.  Thus the product
$\rho_1 \rho_2 \rho_3$ is the overall intrinsic parity of an operator.  

The states of three spin 1/2 objects span a space of three IRs: 
a two-dimensional mixed antisymmetric IR ({\it MA}); 
a two-dimensional mixed symmetric IR ({\it MS}); and 
a four-dimensional totally symmetric IR ({\it S}).  
The {\it MA} and {\it MS} 
are $G_1$ IRs 
and {\it S} is $H$ IR.  
Three $\rho$-spins can be categorized into the same combinations.  
Ten direct products of $\rho$-spin and $s$-spin that 
span the positive-parity 
IRs of the group and 
yield the {\it MA} Young tableaux of Dirac indices 
are listed in Table \ref{table1}.
\begin{table}[pt]
\begin{center}
{\footnotesize
\caption{\footnotesize $I=I_z={1\over2}$, local, positive-parity operators.}
\label{table1}
\renewcommand{\arraystretch}{1.2} 
\begin{tabular}{lc}
\hline
$\overline\Psi^{IR^{\cal{P}},k}_{S, S_z} $&
        $ \overline N^{\rho_1\rho_2\rho_3}_{s_1 s_2 s_3}~~ \rm{notation} $ \\
\hline
$\overline\Psi^{H^+}_{3/2, 3/2}$&$ \overline N^{+--}_{+++}   $ \\
$\overline\Psi^{H^+}_{3/2, 1/2}$&$3^{-1/2}(\overline N^{+--}_{++-} + \overline N^{+--}_{+-+} + \overline N^{+--}_{-++}) $ \\
$\overline\Psi^{H^+}_{3/2, -1/2}$&$3^{-1/2}(\overline N^{+--}_{+--} + \overline N^{+--}_{-+-} + \overline N^{+--}_{--+}) $ \\
$\overline\Psi^{H^+}_{3/2, -3/2}$&$\overline N^{+--}_{---}   $ \\
\hline
$\overline\Psi^{G_1^+,1}_{1/2, 1/2}$&$\overline N^{+++}_{+-+}  $ \\
$\overline\Psi^{G_1^+,1}_{1/2, -1/2}$&$\overline N^{+++}_{+--}   $ \\
\hline
$\overline\Psi^{G_1^+,2}_{1/2, 1/2}$&$3^{-1/2}(\overline N^{+--}_{+-+} + \overline N^{-+-}_{+-+} + \overline N^{--+}_{+-+}) $ \\
$\overline\Psi^{G_1^+,2}_{1/2, -1/2}$&$3^{-1/2}(\overline N^{+--}_{+--} + \overline N^{-+-}_{+--} + \overline N^{--+}_{+--}) $ \\
\hline
$\overline\Psi^{G_1^+,3}_{1/2, 1/2}$&
     $6^{-1/2}(2\overline N^{+--}_{++-}-\overline N^{+--}_{+-+}-\overline N^{+--}_{-++}) $ \\
$\overline\Psi^{G_1^+,3}_{1/2, -1/2}$&
     $6^{-1/2}(\overline N^{+--}_{+--}+\overline N^{+--}_{-+-}-2\overline N^{+--}_{--+}) $ \\
\hline
\end{tabular}
} 
\end{center}
\vspace{-0.4cm}
\end{table}
Ten negative-parity IR operators 
are obtained by reversing $\rho_1, \rho_2$ and $\rho_3$.  
There are three embeddings of $G_1$.  
Though there exists no group element that transforms an 
operator from $(G_1, i)$ to $(G_1, j\neq i)$,
in lattice QCD simulations different embeddings should overlap because 
they share the same quantum numbers.
The optimal linear combination of operators from different embeddings 
is found by the variational technique for 
ground and excited states~\cite{Sato03,Subhasish04}.

For isospin 3/2 delta operators, the primitive local operator is chosen to be,
\begin{equation}
\overline{\Delta^{++}}^{\rho_1 \rho_2 \rho_3}_{s_1 s_2 s_3} = \epsilon_{abc} 
\overline u^{\rho_1 a}_{s_1}
\overline u^{\rho_2 b}_{s_2}
\overline u^{\rho_3 c}_{s_3} .
\nonumber
\end{equation}
Ten positive-parity 
operators for $\overline{\Delta^{++}}$ (two $H$ IRs and one $G_1$ IR) are given 
in Table \ref{table2}.  
Ten negative-parity IR operators are obtained by 
reversing $\rho_1, \rho_2$ and $\rho_3$.  
\begin{table}[ht]
\begin{center}
{\footnotesize
\caption{\footnotesize $I=I_z={3\over2}$, local, positive-parity operators.
Only $ ^{\rho_1 \rho_2 \rho_3}_{s_1 s_2 s_3}$ indices are listed.  
Those indices in curly brackets
represent the sum over all cyclic permutations of indices.}
\label{table2}
\renewcommand{\arraystretch}{1.2} 
\begin{tabular}{lc}
\hline
$\overline \Psi^{IR^{\cal{P}},k}_{S, S_z} $&
           $ \overline \Delta^{\rho_1\rho_2\rho_3}_{s_1 s_2 s_3}$ notation \\
\hline
$\overline \Psi^{H^+, 1}_{3/2, 3/2} $&$ ^{+++}_{+++} $ \\
$\overline \Psi^{H^+, 1}_{3/2, 1/2} $&$ 3^{-1/2}\{ ^{+++}_{++-} \} $ \\
$\overline \Psi^{H^+, 1}_{3/2,-1/2} $&$ 3^{-1/2}\{ ^{+++}_{+--} \} $ \\
$\overline \Psi^{H^+, 1}_{3/2,-3/2} $&$ ^{+++}_{---} $ \\
\hline
$\overline \Psi^{H^+, 2}_{3/2, 3/2} $&$ 3^{-1/2}\{ ^{+--}_{+++} \} $ \\
$\overline \Psi^{H^+, 2}_{3/2, 1/2} $&$ 
3^{-1}(\{ ^{+--}_{++-} \} + \{ ^{+--}_{+-+} \} + \{ ^{+--}_{-++} \} )$ \\
$\overline \Psi^{H^+, 2}_{3/2, 1/2} $&$ 
3^{-1}(\{ ^{+--}_{+--}\} + \{ ^{+--}_{-+-}\} + \{ ^{+--}_{--+}\} )$ \\
$\overline \Psi^{H^+, 2}_{3/2,-3/2} $&$ 3^{-1/2}\{ ^{+--}_{---}\} $ \\
\hline
$\overline \Psi^{G_1^+}_{1/2, 1/2} $&
       $ 2^{-1}( ^{+--}_{+-+} -  ^{+--}_{-++} -  ^{-+-}_{+-+} +  ^{-+-}_{-++})$ \\
$\overline \Psi^{G_1^+}_{1/2,-1/2} $&
       $ 2^{-1}( ^{+--}_{+--} -  ^{+--}_{-+-} -  ^{-+-}_{+--} +  ^{-+-}_{-+-})$ \\
\hline
\end{tabular}
} 
\end{center}
\vspace{-0.4cm}
\end{table}

Table \ref{table1} can be used for other baryon channels that have {\it MA} flavor 
content, {\it e.g.}, 
$\Lambda  \sim uds$.  
Similarly,
Table \ref{table2} can be used for baryons with {\it S} flavor content, {\it e.g.}, 
$\Omega   \sim sss$, 
$\Sigma^+ \sim uus$, and 
$\Xi^0    \sim uss$.

\section{STRAIGHT-LINK OPERATORS}
Here, we consider the simplest nonlocal operator---the third quark displaced along
a straight path from the other two, 
\begin{eqnarray}
\overline O_i ^{(n)} \equiv \epsilon_{abc} 
\overline q_{1\mu_1}^a ({\bf x}) 
\overline q_{2\mu_2}^b ({\bf x}) 
[f(U)
\overline q_{3\mu_3} ({\bf x}+na{\bf \hat e}_i) ]^c,
\nonumber
\end{eqnarray}
where $n$ is an integer specifying the number of sites displaced, 
and $f(U)$ is a product of gauge links connecting the third quark to point ${\bf x}$.
Dirac indices are suppressed in $\overline O_i ^{(n)}$.  
There are six possible link directions for 
$\overline O_i ^{(n)}$ that 
transform amongst themselves according to the octahedral group.  
They are reduced into three IRs: $A_1$, $T_1$, and $E$ 
with dimensions 1, 3, and 2, respectively.  
The $A_1$-link operator is the sum of all six $\overline O_i$.  Because its spatial 
distribution 
is cubically symmetric, 
IR operators for $A_1$-link can be obtained directly from Table \ref{table1}
or Table \ref{table2}.
We choose $T_1$- and $E$-link operators to be discretized versions of 
a spherical basis, 
following~\cite{Wingate95,Lacock96}:
\begin{eqnarray}
D_0 \overline B   &\equiv& (-i/2a)(\overline O_z - \overline O_{-z});\nonumber \\
D_\pm \overline B &\equiv& (\pm i/ 2\sqrt{2}a)[(\overline O_x \! - \! \overline O_{\! -x}) 
	\! \pm \! i(\overline O_y \! - \! \overline O_{\! -y})];\nonumber \\
E_0 \overline B   &\equiv& (1/ \sqrt{6} a^2)
	[ 2(\overline O_z + \overline O_{\! -z}) - (\overline O_x + \overline O_{\! -x}) \nonumber \\
		  & & - (\overline O_y + \overline O_{-y})];\nonumber \\
E_2 \overline B   &\equiv& (1/ \sqrt{2} a^2) [(\overline O_x + \overline O_{\! -x}) - 
	(\overline O_y + \overline O_{\! -y})], \nonumber
\end{eqnarray}
where $D_m$($E_m$) acting on the third quark of baryon source 
$\overline B$ are $T_1$-($E$-) link operators.
\begin{table}[h]
\caption{\footnotesize Straight-link, P-wave operators.}
\label{table3}
\renewcommand{\arraystretch}{1.3} 
{\footnotesize
\begin{tabular}{cc}
\hline
$H^{\mp}$ operators ($J=3/2$)  &  $J_z$ \\
\hline
$D_+ \overline \Psi^{G_1^\pm, k}_{{1\over2},{1\over2}}$   &
	$+{3\over2}$ \\
${1\over \sqrt{3}} D_+ \overline \Psi^{G_1^\pm, k}_{{1\over2},-{1\over2}}
	+  \sqrt{2\over 3}D_0 \overline \Psi^{G_1^\pm, k}_{{1\over2},{1\over2}}$   &
	$+{1\over2}$ \\
$\sqrt{2\over 3}D_0 \overline \Psi^{G_1^\pm, k}_{{1\over2},-{1\over2}}
        + {1\over \sqrt{3}} D_- \overline \Psi^{G_1^\pm, k}_{{1\over2},{1\over2}}$  &
	$-{1\over2}$ \\
$D_- \overline \Psi^{G_1^\pm, k}_{{1\over2},-{1\over2}}$   &
	$-{3\over2}$ \\
\hline
$ \sqrt{3\over 5}D_0 \overline \Psi^{H^\pm, k}_{{3\over2},{3\over2}}
        -{\sqrt{2 \over 5}} D_+ \overline \Psi^{H^\pm, k}_{{3\over2},{1\over2}}$  &
	$+{3\over2}$ \\
$ \sqrt{2\over 5}D_- \overline \Psi^{H^\pm, k}_{{3\over2},{3\over2}}
	+ {1 \over \sqrt{15}} D_0 \overline \Psi^{H^\pm, k}_{{3\over2},{1\over2}}
        - {\sqrt{8 \over 15}} D_+ \overline \Psi^{H^\pm, k}_{{3\over2},-{1\over2}} $ &
	$+{1\over2}$ \\
$ \sqrt{8\over 15}D_- \overline \Psi^{H^\pm, k}_{{3\over2},{1\over2}}
	- {1 \over \sqrt{15}} D_0 \overline \Psi^{H^\pm, k}_{{3\over2},-{1\over2}}
        - {\sqrt{2 \over 5}} D_+ \overline \Psi^{H^\pm, k}_{{3\over2},-{3\over2}}$  &
	$-{1\over2}$ \\
${\sqrt{2 \over 5}} D_- \overline \Psi^{H^\pm, k}_{{3\over2},-{1\over2}}
	-  \sqrt{3\over 5}D_0 \overline \Psi^{H^\pm, k}_{{3\over2},-{3\over2}}$   &
	$-{3\over2}$ \\
\hline
$H^{\mp}$ operators (row 1 and 4 have mixed $J$, & \\
	row 2 and 3 have $J=5/2$) & \\
\hline
${1\over \sqrt{10}} D_+ \overline \Psi^{H^\pm, k}_{{3\over2},{1\over2}}
	+  \sqrt{5\over 6}D_- \overline \Psi^{H^\pm, k}_{{3\over2},-{3\over2}}
	+ {1 \over \sqrt{15}} D_0 \overline \Psi^{H^\pm, k}_{{3\over2},{3\over2}}$  & 
\\
$ {1\over \sqrt{10}}D_- \overline \Psi^{H^\pm, k}_{{3\over2},{3\over2}}
	+ \sqrt{3\over 5} D_0 \overline \Psi^{H^\pm, k}_{{3\over2},{1\over2}}
        + {\sqrt{3 \over 10}} D_+ \overline \Psi^{H^\pm, k}_{{3\over2},-{1\over2}} $ & $+{1\over2}$
	   \\
$ \sqrt{3\over 10}D_- \overline \Psi^{H^\pm, k}_{{3\over2},{1\over2}}
	+ \sqrt{3\over 5} D_0 \overline \Psi^{H^\pm, k}_{{3\over2},-{1\over2}}
	+ {1\over \sqrt{10}} D_+ \overline \Psi^{H^\pm, k}_{{3\over2},-{3\over2}}$  & $-{1\over2}$
	   \\
${\sqrt{5 \over 6}} D_+ \overline \Psi^{H^\pm, k}_{{3\over2},{3\over2}}
	+  {1\over \sqrt{10}}D_- \overline \Psi^{H^\pm, k}_{{3\over2},-{1\over2}}
	+ {1 \over \sqrt{15}} D_0 \overline \Psi^{H^\pm, k}_{{3\over2},-{3\over2}}$  & 
\\
\hline
$G_1^\mp$ operators ($J=1/2$) & \\
\hline
$\sqrt{2\over 3} D_+ \overline \Psi^{G_1^\pm, k}_{{1\over2},-{1\over2}}
	- {1\over \sqrt{3}}D_0 \overline \Psi^{G_1^\pm, k}_{{1\over2},{1\over2}}$   &
	$+{1\over2}$ \\
$ {1\over \sqrt{3}}D_0 \overline \Psi^{G_1^\pm, k}_{{1\over2},-{1\over2}}
	- \sqrt{2\over 3} D_- \overline \Psi^{G_1^\pm, k}_{{1\over2},{1\over2}}$   &
	$-{1\over2}$ \\
\hline
${1\over \sqrt{2}} D_- \overline \Psi^{H^\pm, k}_{{3\over2},{3\over2}}
	-  {1\over \sqrt{3}}D_0 \overline \Psi^{H^\pm, k}_{{3\over2},{1\over2}}
	+  {1\over \sqrt{6}}D_+ \overline \Psi^{H^\pm, k}_{{3\over2},-{1\over2}}$  &
	$+{1\over2}$ \\
$ {1\over \sqrt{6}}D_- \overline \Psi^{H^\pm, k}_{{3\over2},{1\over2}}
	-  {1\over \sqrt{3}}D_0 \overline \Psi^{H^\pm, k}_{{3\over2},-{1\over2}}
	+ {1\over \sqrt{2}} D_+ \overline \Psi^{H^\pm, k}_{{3\over2},-{3\over2}}$  &
	$-{1\over2}$ \\
\hline
$G_2^\mp$ operators (mixed $J$) & \\
\hline
$ {1\over \sqrt{2}}D_- \overline \Psi^{H^\pm, k}_{{3\over2},-{1\over2}}
	+  {1\over \sqrt{3}}D_0 \overline \Psi^{H^\pm, k}_{{3\over2},-{3\over2}}
	- {1\over \sqrt{6}} D_+ \overline \Psi^{H^\pm, k}_{{3\over2},{3\over2}}$  & 
\\
${1\over \sqrt{6}} D_- \overline \Psi^{H^\pm, k}_{{3\over2},-{3\over2}}
	-  {1\over \sqrt{3}}D_0 \overline \Psi^{H^\pm, k}_{{3\over2},{3\over2}}
	-  {1\over \sqrt{2}}D_+ \overline \Psi^{H^\pm, k}_{{3\over2},{1\over2}}$  &
\\
\hline
\end{tabular}
}
\vspace{-0.4cm}
\end{table}
\begin{table}[h]
\begin{center}
\caption{\footnotesize Straight-link, D-wave operators.}
\label{table4}
\renewcommand{\arraystretch}{1.6} 
{\footnotesize
\begin{tabular}{ccc}
\hline
 & $H^{\pm}$ operators (mixed $J$)  & \\
\hline
 & $E_2\overline\Psi^{G_1^{\pm},k}_{1/2,-1/2}$ & \\
 & $E_0\overline\Psi^{G_1^{\pm},k}_{1/2,1/2}$  & \\
 & $E_0\overline\Psi^{G_1^{\pm},k}_{1/2,-1/2}$ & \\
 & $E_2\overline\Psi^{G_1^{\pm},k}_{1/2,1/2}$  & \\
\hline
 & $E_0\overline\Psi^{H^{\pm},k}_{3/2,3/2} + E_2\overline\Psi^{H^{\pm},k}_{3/2,-1/2}$ & \\
 & $E_0\overline\Psi^{H^{\pm},k}_{3/2,1/2} - E_2\overline\Psi^{H^{\pm},k}_{3/2,-3/2}$ & \\
 & $E_0\overline\Psi^{H^{\pm},k}_{3/2,-1/2} - E_2\overline\Psi^{H^{\pm},k}_{3/2,3/2}$ & \\
 & $E_0\overline\Psi^{H^{\pm},k}_{3/2,-3/2} + E_2\overline\Psi^{H^{\pm},k}_{3/2,1/2}$ & \\
\hline
 & $G_1^{\pm}$ operators  (mixed $J$) & \\
\hline
 & $E_0\overline\Psi^{H^{\pm},k}_{3/2,1/2} + E_2\overline\Psi^{H^{\pm},k}_{3/2,-3/2}$ & \\
 & $E_0\overline\Psi^{H^{\pm},k}_{3/2,-1/2} + E_2\overline\Psi^{H^{\pm},k}_{3/2,3/2}$ & \\
\hline
 & $G_2^{\pm}$ operators  (mixed $J$) & \\
\hline
 & $-E_0\overline\Psi^{H^{\pm},k}_{3/2,3/2} + E_2\overline\Psi^{H^{\pm},k}_{3/2,-1/2}$ & \\
 & $-E_0\overline\Psi^{H^{\pm},k}_{3/2,-3/2}+ E_2\overline\Psi^{H^{\pm},k}_{3/2,1/2}$  & \\
\hline
\end{tabular}
}
\end{center}
\vspace{-0.5cm}
\end{table}
The $T_1$-($E$-) link operators have odd(even) spatial parity.  
The decomposition of vectorial IR $\otimes$ spinorial IR is, 
\begin{eqnarray}
T_1 \otimes \,H\, &=& H \oplus H   \oplus G_1 \oplus G_2, \nonumber \\
T_1 \otimes G_1   &=& H \oplus G_1,                        \nonumber \\
\,E\, \otimes \,H\, &=& H \oplus G_1 \oplus G_2,            \nonumber \\
\,E\, \otimes G_1 &=& H.
\label{decomposition}
\end{eqnarray}
Tables \ref{table3} and \ref{table4} show all IR operators formed from 
straight-link constructions according to Eq.(\ref{decomposition}).  
The local operators $\overline\Psi^{IR^{\cal P}, k}_{S, S_z}$ are selected from 
Table \ref{table1} or Table \ref{table2}.
The $D_m$ straight-link operators are analogous to the discretized spherical 
harmonics $Y_{1m}$, hence 
we list the $J_z$ values in Table \ref{table3} 
for operators having Clebsch-Gordan coefficients.  
Straight-link operators include $G_2$ operators, which 
couple to spin states with $J\geq 5/2$.

To verify that the operators in Table \ref{table1}-\ref{table3} satisfy the 
appropriate orthogonality relations, we have performed a test using 
287 quenched, $16^3 \times 64$ lattices with the Wilson fermion action.  
We have confirmed that the correlation functions vanish within one jackknife 
error for all timeslices when the source and sink operators belong to 
different IRs ($G_1, G_2, H$), have different parities, or belong to 
different rows of the same IR.
 
This work was supported by the U.S. National Science Foundation under
Awards PHY-0099450 and PHY-0300065, and by the U.S. Department of Energy 
under contracts DE-AC05-84ER40150 and DE-FG02-93ER-40762.


%
\end{document}